# PHYSICS AT AN $e^-e^-$ FACILITY

FRANK CUYPERS
cuypers@pss058.psi.ch
*Max-Planck-Institut für Physik, Föhringer Ring 6, D-80805 München, Germany*

We review some of the reactions which can be studied in the $e^-e^-$ mode of a linear collider and may reveal aspects of physics beyond the realm of the standard model. The complementarity to $e^+e^-$ scattering is stressed.

## 1 Introduction

There are several important characteristics which differentiate $e^-e^-$ from $e^+e^-$ collisions, and justify considering both options on the same footing when it comes to designing the linear colliders of the next generation:

- The $e^-e^-$ environment is much cleaner because there is much less standard model activity. In particular, since QCD enters the game only at much higher orders and is always associated with large missing energy, there are no systematic errors due to the possible misidentification of electrons and pions.

- Current electron guns can already produce beams with polarizations exceeding 70%. There is no doubt that even further improvements are due. It not clear, however, whether polarized positron beams may ever be obtained at all. The $e^-e^-$ collisions therefore offer the possibility of polarizing both initial states and to perform three independent experiments.

- The $e^-e^-$ initial state is not only doubly charged, but also carries a finite lepton number. This allows to explore fermion number or flavour violating interactions, which are difficult to access in conventional $e^+e^-$ annihilations.

These features make the $e^-e^-$ linear collider mode a powerful tool for probing phenomena outside the scope of the standard model [1]. We review here several areas where $e^-e^-$ scattering can provide informations which are at least complementary to those that can be gathered in the $e^+e^-$ or other operating modes.

## 2 $Z'$ Bosons[2]

There are many extensions of the standard model which predict the existence of extra neutral vector bosons[3]. Most assume lepton universality, so that the generic lagrangian describing the interaction of a heavy neutral vector boson $Z'$ with the charged leptons can be written

$$L = e\ \bar{\psi}_\ell \gamma^\mu (v_{Z'} + a_{Z'}\gamma_5)\, \psi_\ell\ Z'_\mu\ , \tag{1}$$

where $v_{Z'}$ and $a_{Z'}$ are the vector and axial couplings. This interaction mediates both $e^+e^-$ annihilation and $e^-e^-$ scattering, as depicted in Fig 1.

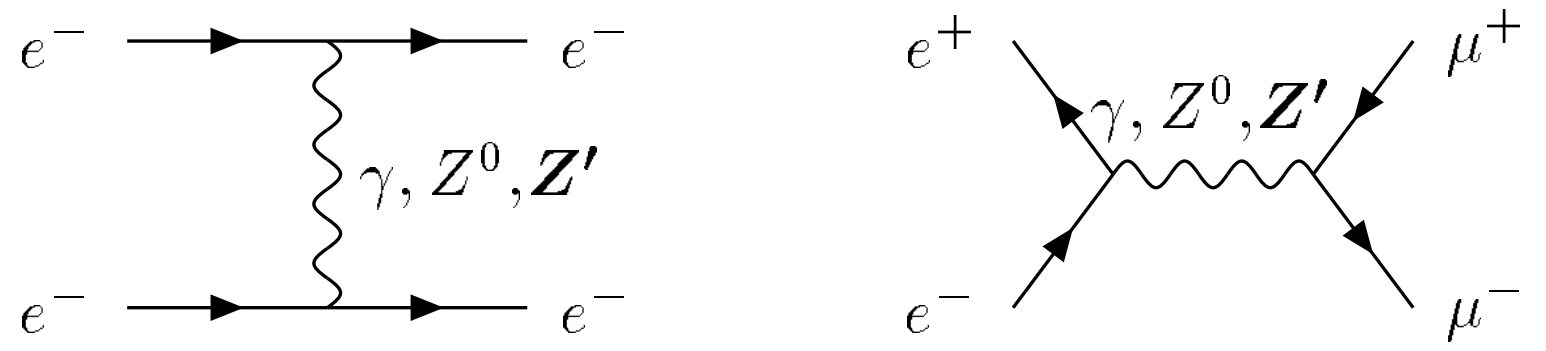

Figure 1: Lowest order Feynman graphs with a $Z'$ exchange in $e^-e^-$ and $e^+e^-$ collisions.

Of course, if the collider energy is sufficient to sit on the $Z'$ resonance, the $e^+e^-$ option cannot be beaten. However, if the $Z'$ mass exceeds the collider energy by as little as about 25% the study of angular correlations in Møller scattering turns starts providing stronger evidence for the existence of a $Z'$ than the $e^+e^- \to \mu^+\mu^-$ reaction. If a $Z'$ is discovered, both experiments provide complementary bounds on the values of the two couplings $v_{Z'}$ and $a_{Z'}$, as can be observed in Fig. 2.

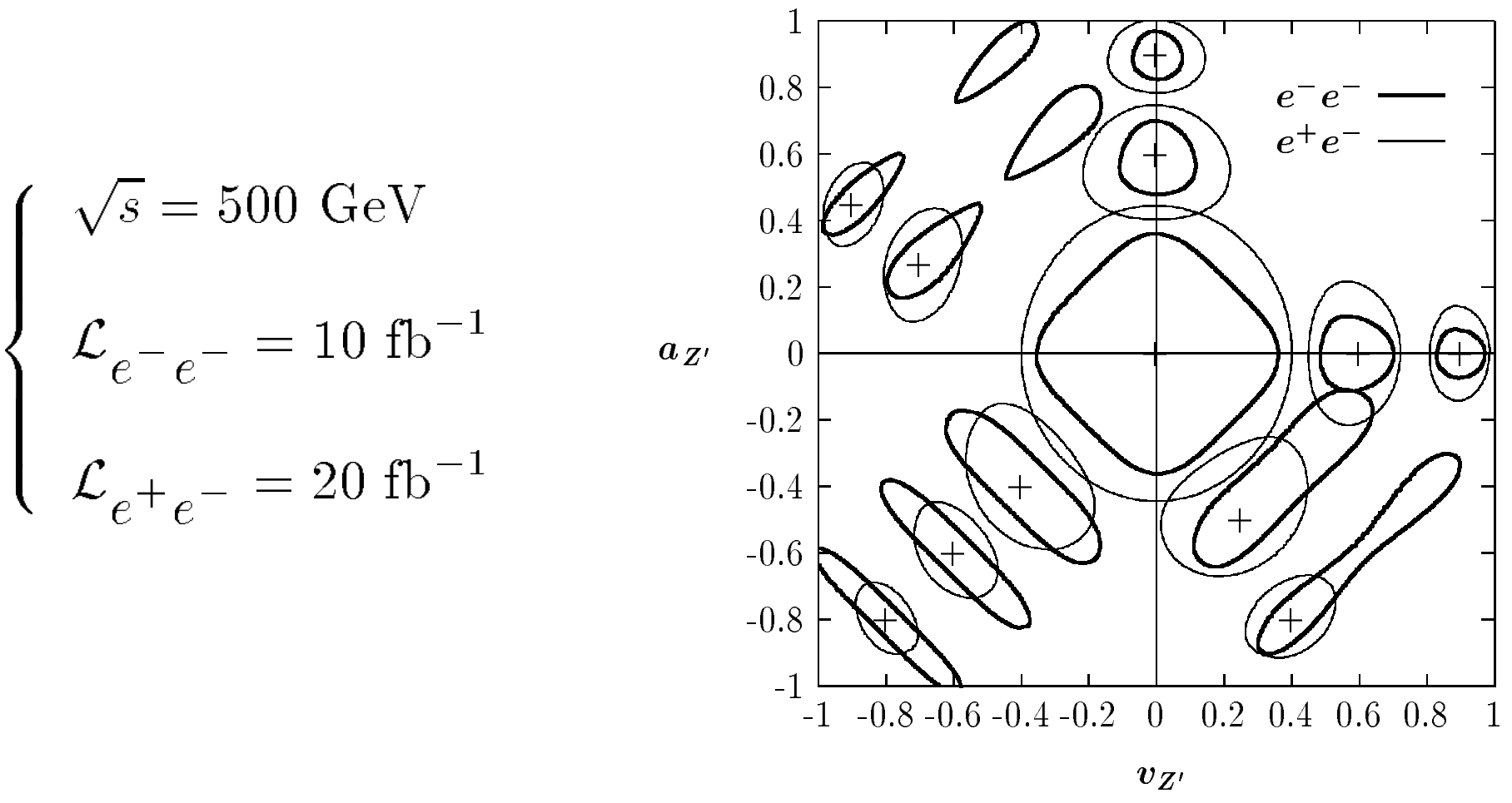

Figure 2: Contours of resolvability at 95% confidence of the $Z'$ couplings around several possible true values marked with a '+'.

## 3 Dileptons [4]

An important class of grand unified schemes predicts the existence of doubly charged vector gauge fields, which couple universally to leptons. These dileptons would show up as striking resonances in $e^-e^-$ scattering, as depicted in Fig. 3.

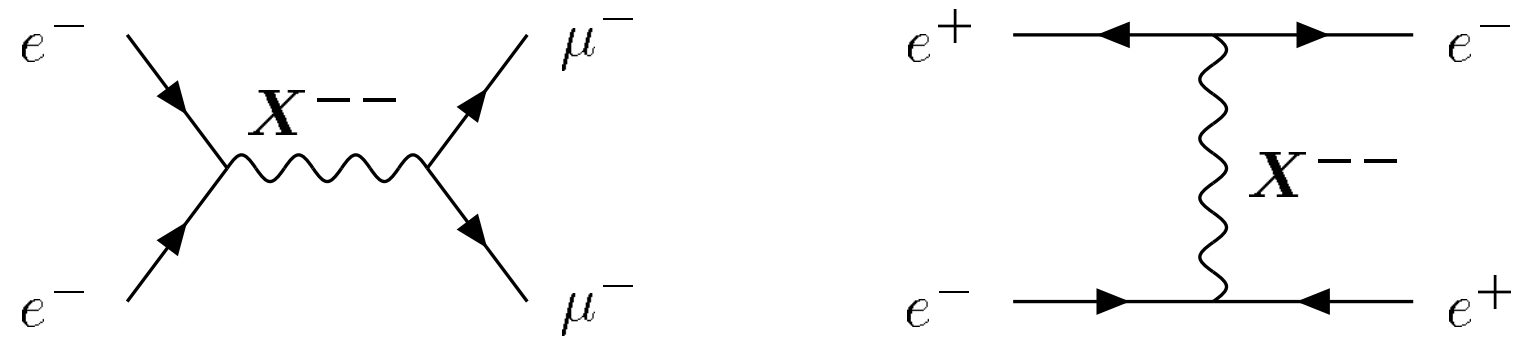

Figure 3: Lowest order Feynman diagram describing a dilepton exchange in $e^-e^-$ and $e^+e^-$ collisions.

The situation here is just opposite from $Z'$ production, in the sense that below threshold it is now Bhabha scattering which provides better discovery limits. Still, as in the $Z'$ case, one may anticipate that both reactions are complementary when it comes to study the couplings of an off-shell dilepton. Of course, once the collider energy comes close to the dilepton resonance, the $e^-e^-$ option is best.

## 4 Leptoquarks [5]

Leptoquarks are predicted by a large number of extensions of the standard model and can appear in many combinations of several quantum numbers [6]. One of these quantum numbers is the fermion number $F = 0, 2$ carried by the leptoquark. It turns out that other experiments are not directly sensitive to $F$ (except for electron-(anti)proton collisions at LEP-LHC, if this option is ever turned on). Therefore, even if a leptoquark is discovered at an $e^+e^-$ or another facility, its true nature may remain hidden.

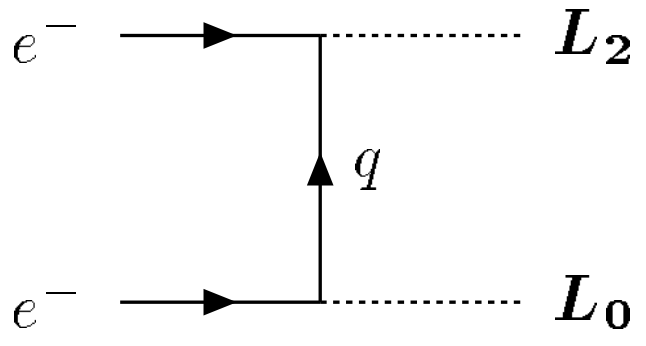

Figure 4: Typical lowest order Feynman diagram describing the production of two leptoquarks (scalar or vector or both) in $e^-e^-$ collisions.

This is where $e^-e^-$ collisions become interesting, because to lowest order they can only produce a pair of leptoquarks, one with fermion number $F = 0$ and the other with $F = 2$, as depicted in Fig. 4. Therefore, the observation of such events would demonstrate the simultaneous existence of these two states. Similarly, the non-observation of this mechanism would impose strong bounds on extensions of the standard model.

To estimate the discovery potential, Fig. 5 shows in the $(m, \lambda/e)$ plane the $\sigma = 1$ fb curve for one of the five possible types of reactions. Assuming the produced leptoquarks have the same mass, the following general discovery scaling law applies

$$\frac{\lambda}{e} \quad = \quad 0.35 \quad \sqrt{m/\mathrm{TeV}} \quad \left(\frac{n}{A\ \mathcal{L}/\mathrm{fb}^{-1}}\right)^{1/4} \qquad\qquad m \leq .43\sqrt{s}\ , \qquad (2)$$

where $e$ is the charge of the electron, $\lambda$ is the geometric mean of the two leptoquarks' couplings, $m$ is their common mass, $n$ is the required number of events, $\mathcal{L}$ is the available luminosity and $A$ is a number ranging between 1 and 24, according to which types of leptoquarks are produced.

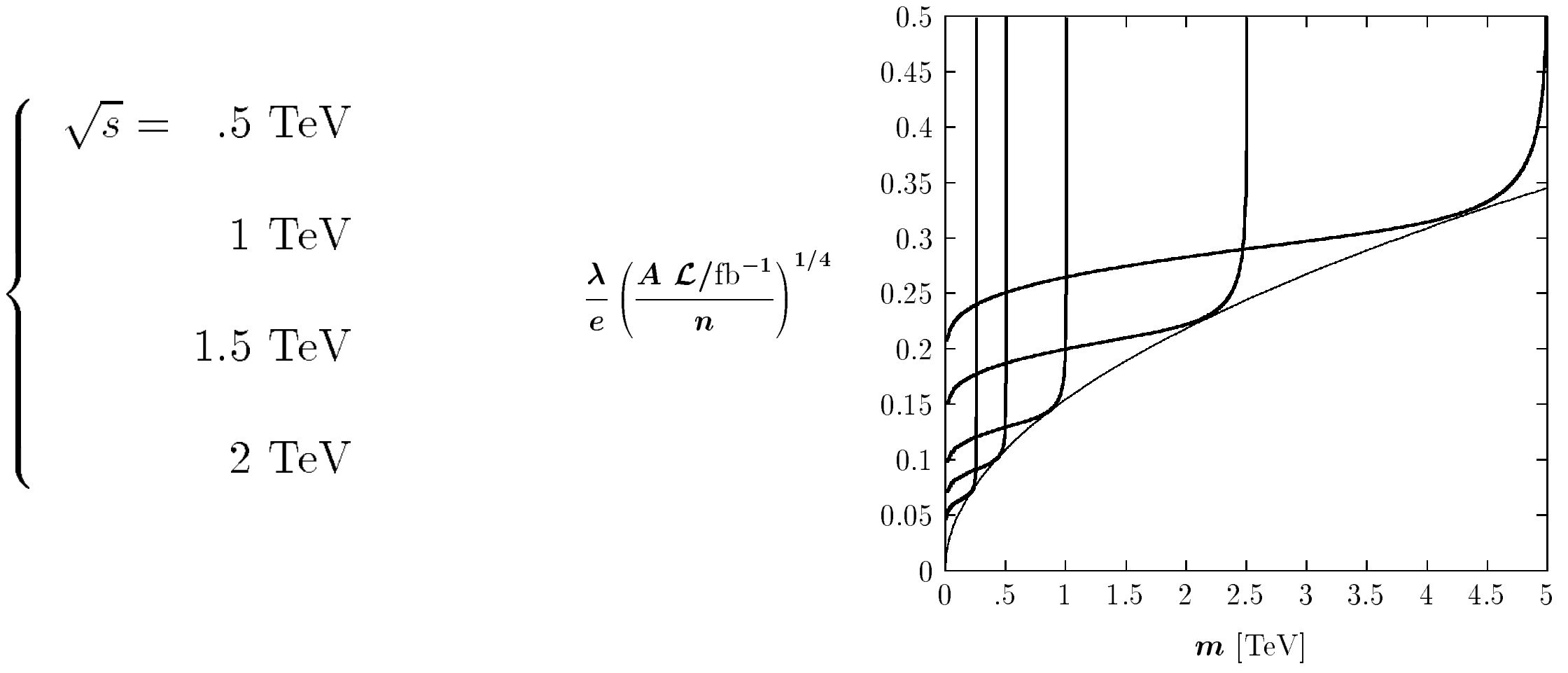

Figure 5: The parameter space below the curves cannot be explored in $e^-e^-$ scattering. The thinner osculating parabola is given by Eq. (2).

## 5 Majorana Neutrinos [7]

In the presence of heavy Majorana neutrinos, $W^-$ pairs can be produced in $e^-e^-$ scattering, as depicted in Fig. 6. Low energy experiments severely constrain the masses and mixings of these states, so that with 500 GeV and 10 $\mathrm{fb}^{-1}$ no more than about 100 such events are expected [8]. In principle, a single occurrence of this fermion number violating transition would suffice to establish a departure from the standard model.

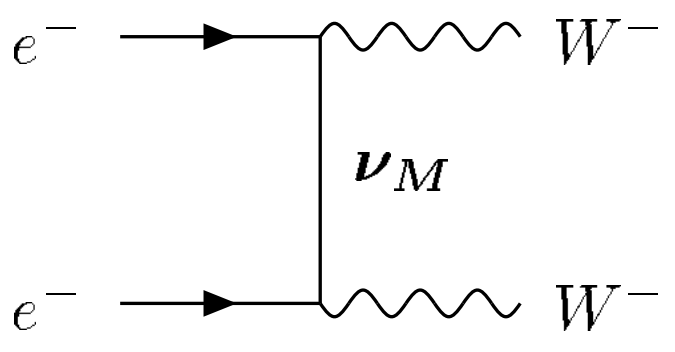

Figure 6: Lowest order Feynman diagram describing $W^-$ pair-production via Majorana neutrino exchange in $e^-e^-$ collisions.

However, the higher order electroweak reactions

$$\begin{aligned} e^-e^- &\to W^-W^-\nu\nu & (3)\\ &\to W^-Z^0\nu e^- & (4)\\ &\to Z^0Z^0e^-e^- & (5)\\ &\to W^+W^-e^-e^- , & (6)\end{aligned}$$

where the $Z^0$'s decay invisibly or hadronically, may very well mimic the exotic reaction. Indeed, most of the primary electrons in reactions (4–6) disappear into the beam-pipe [9]. As can be gathered from Fig. 7, the cross sections are substantial, so that the leptonic decays of the pair-produced signal $W^-$'s are unlikely to emerge from the backgrounds. One therefore has to concentrate on the hadronic decays, but then also the $Z^0$'s may be mistaken for $W^-$'s if the jet invariant mass resolution is insufficient.

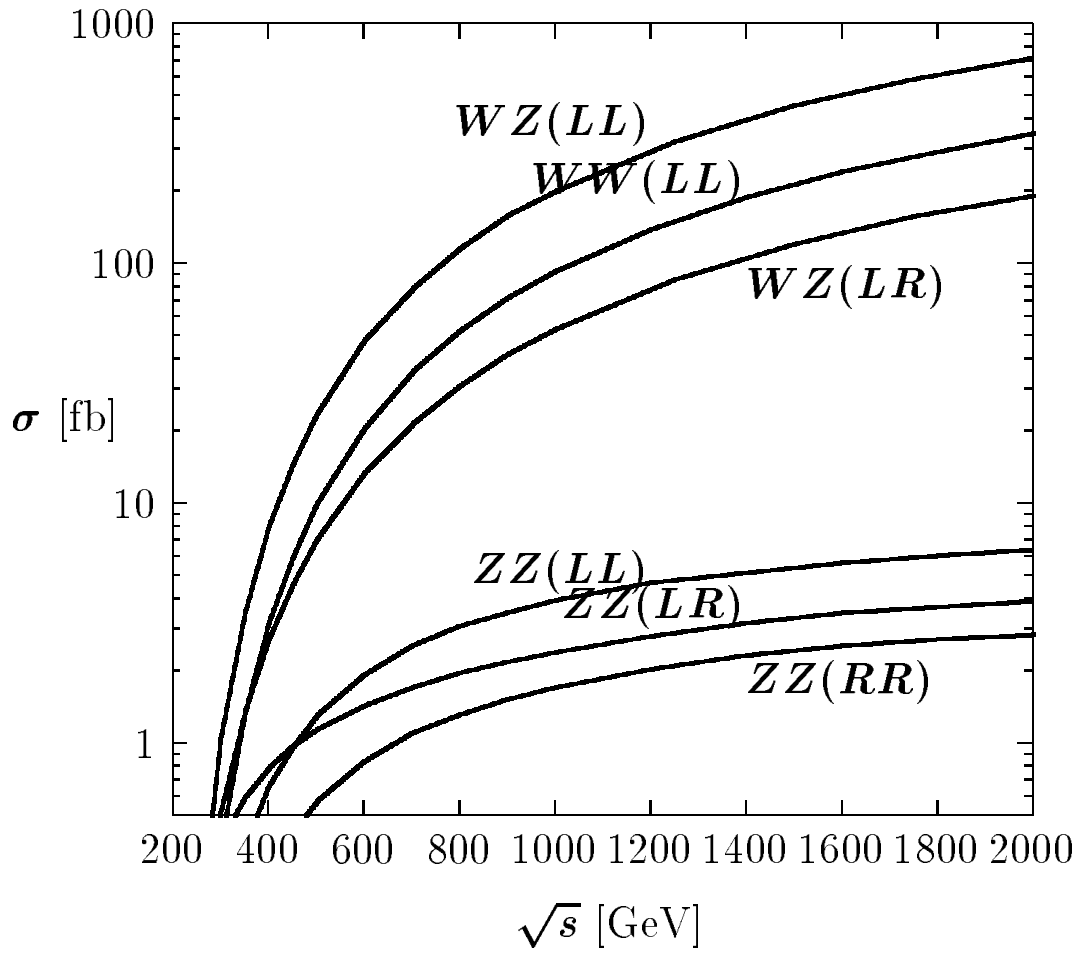

Figure 7: Total cross sections for gauge boson pair-production in polarized $e^-e^-$ scattering.

Nevertheless, the backgrounds (3–6) can easily be removed by considering their total hadronic energy or transverse momentum distributions. These are depicted in Fig. 8 for reaction (4) and show basically no overlap with the (smoothed) Dirac distributions centered at $\sqrt{s}$ and 0, which are expected from the 2-body reaction $e^-e^- \to W^-W^-$. Therefore, a Majorana signal cannot be mistaken for any standard model process.

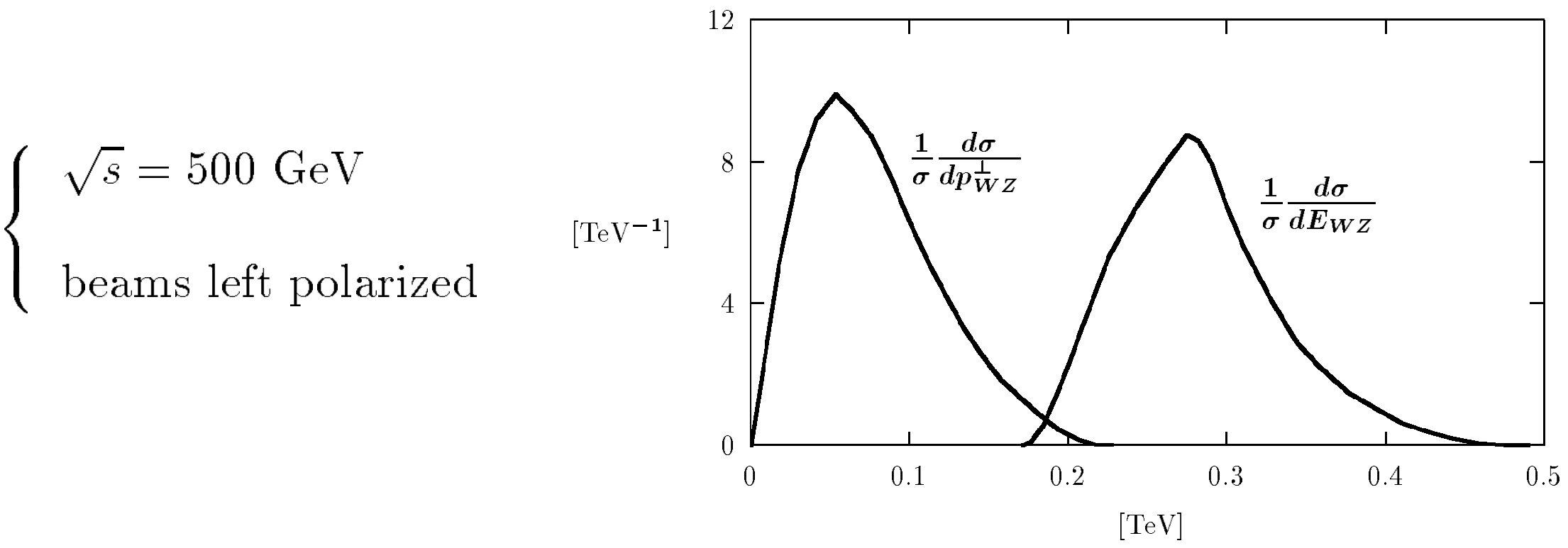

Figure 8: Transverse momentum and energy distributions of the $W^-Z^0$-pair in reaction (4).

## 6 Selectrons [10,11]

Selectrons, the supersymmetric scalar partners of the electrons [12], can be pair-produced in $e^-e^-$ scattering via the exchange of four neutralinos, as depicted in Fig. 9.

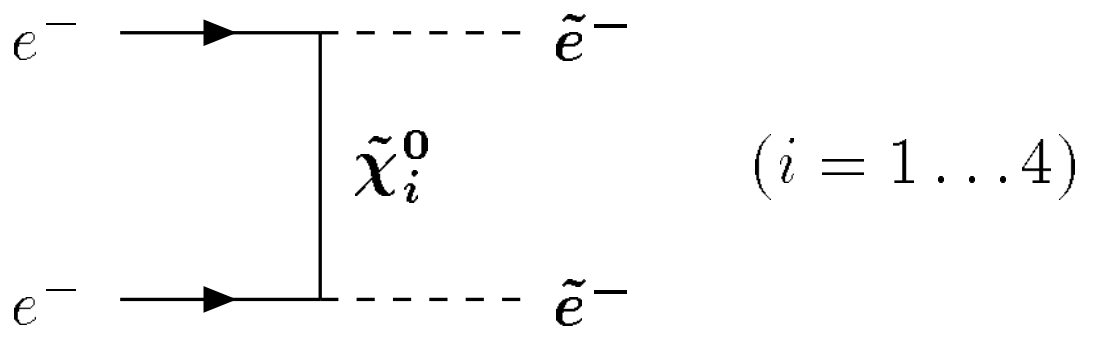

Figure 9: Lowest order Feynman diagram describing selectron production via the exchange of neutralinos in $e^-e^-$ collisions.

This reaction is described elsewhere in these proceedings [13]. Let us just repeat here that it suffers from very little standard model backgrounds and is thus ideally suited for discovering the selectron. Moreover, the background-free environment allows for a very precise kinematical (hence model-independent) measurement of the mass of the lightest neutralino, which in the minimal supersymmetric standard model escapes detection. Similarly, since no hefty energy or transverse momentum cuts are required to separate signal from background, possible cascade decays of the selectron will be clearly observable [11].

## 7 Higgs Bosons [14,15,16]

The hunt for the standard model Higgs boson is certainly one of the most pressing issues in high energy physics today. Also an $e^-e^-$ experiment may provide further information about this mysterious sector via $Z^0$ fusion [14], as depicted in Fig. 10. Here again, the low standard model background is an important asset.

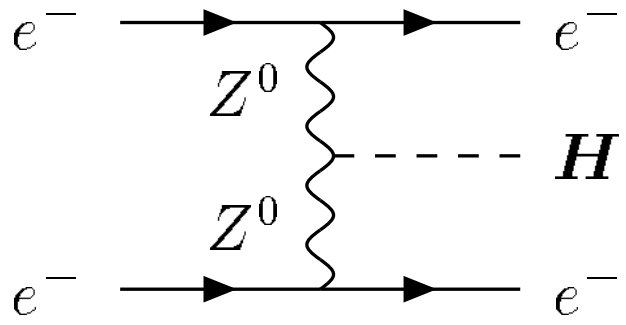

Figure 10: Lowest order Feynman diagram describing the production of the standard model Higgs bosons in $e^-e^-$ collisions.

Similarly, the $W$ fusion mechanism depicted in the first diagram of Fig. 11 can produce two charged Higgs', which typically arise in a two doublet Higgs model [15]. This reaction, which involves the full spectrum of neutral Higgs', is a particularly sensitive probe of the extended Higgs nature.

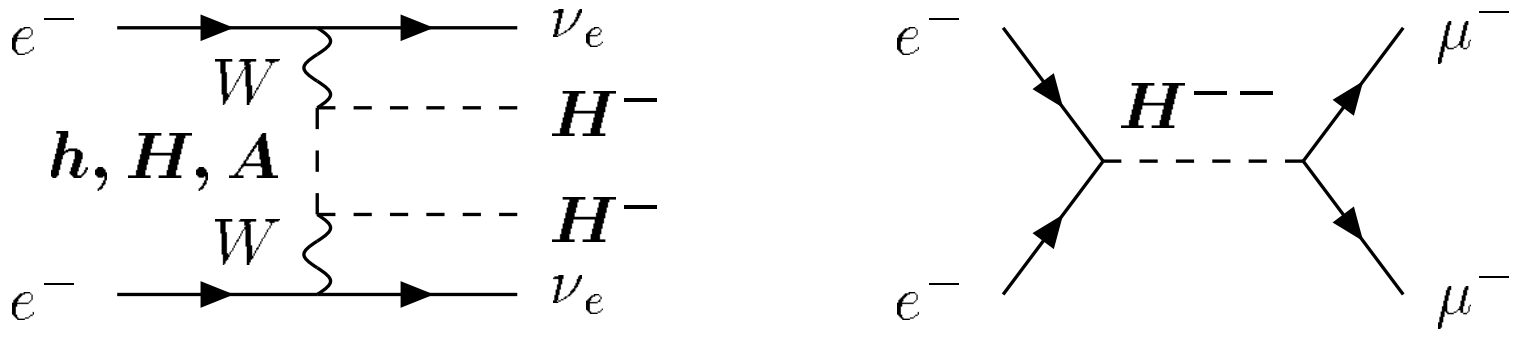

Figure 11: Lowest order Feynman diagrams describing the production of charged Higgs bosons in $e^-e^-$ collisions.

The full potential of $e^-e^-$ collisions, though, is realized in the search for a doubly charged Higgs [16], which could be produced in the $s$-chanel reaction of Fig. 11. The corresponding exotic Higgs representation is not necessarily ruled out by the observation that $\rho = 1$, if its neutral member has no vacuum expectation value. The exploration of this doubly charged resonance by an energy scan would provide most remarkable information about the Higgs sector.

## 8 Anomalous Gauge Couplings [17,18]

Vector boson self-couplings are one of the most eminent consequences of a non-abelian symmetry. Their form is precisely dictated by the gauge principle, so that any deviation may provide crucial information about the new physics lurking at the TeV scale.

To parametrize our ignorance about the latter, it is customary to consider an effective lagrangian of the form [19]

$$L_{\text{eff}}^{WWV} = -ig_V\left[\boldsymbol{g_V^1}\left(W^\dagger_{\alpha\beta}W^\alpha - W^{\dagger\alpha}W_{\alpha\beta}\right)V^\beta + \boldsymbol{\kappa_V}W^\dagger_\alpha W_\beta V^{\alpha\beta} + \frac{\boldsymbol{\lambda_V}}{M_W^2}W^\dagger_{\alpha\beta}W^\beta{}_\sigma V^{\sigma\alpha}\right] \qquad (V=\gamma, Z)\,, \tag{7}$$

where $V_{\alpha\beta} = \partial_\alpha V_\beta - \partial_\beta V_\alpha$ and $W_{\alpha\beta} = \partial_\alpha W_\beta - \partial_\beta W_\alpha$. The standard model prediction is recovered by setting $\kappa_V = g_V^1 = 1$ and $\lambda_V = 0$. Other $C$ and/or $P$ violating anomalous terms can also be added, and there are similar extensions of the quartic part of the lagrangian. These anomalous couplings are determined or constrained within particular models.

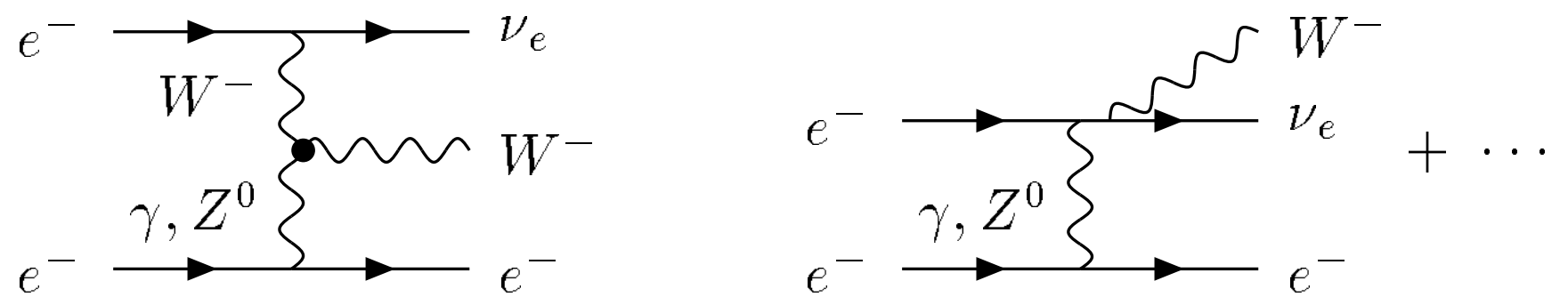

Figure 12: Lowest order Feynman diagram involving a triple gauge vertex and one of its backgrounds in $e^-e^-$ collisions.

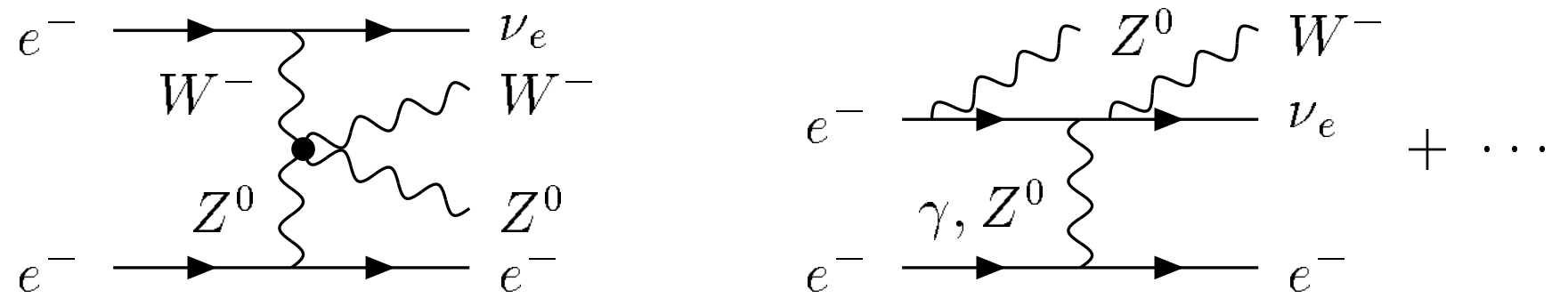

Figure 13: Lowest order Feynman diagram involving a quartic gauge vertex and one of its many backgrounds in the reaction (4).

Anomalous triple [17] or quartic [18] vertices can be probe through $W^-$ production in $e^-e^-$ scattering, as depicted in Figs 12 and 13. In the case of a strongly interacting Higgs sector, for instance, for which the anomalies are severely constrained, these reactions are described with more details elsewhere in these proceedings [20]. Let us only mention here, that in general the use of polarized beams significantly enhances the resolving power, as is clearly demonstrated in Fig. 14. This resolving power is comparable to what can be achieved with a similar analysis and assumptions in other linear collider modes, such as $e^+e^-$, $e^-\gamma$ or $\gamma\gamma$ scattering.

$$\left\{ \begin{array}{l} \sqrt{s} = 500 \text{ GeV} \\ \mathcal{L}_{e^-e^-} = 10 \text{ fb}^{-1} \\ \text{hadronic } W^- \text{ decays only} \end{array} \right.$$

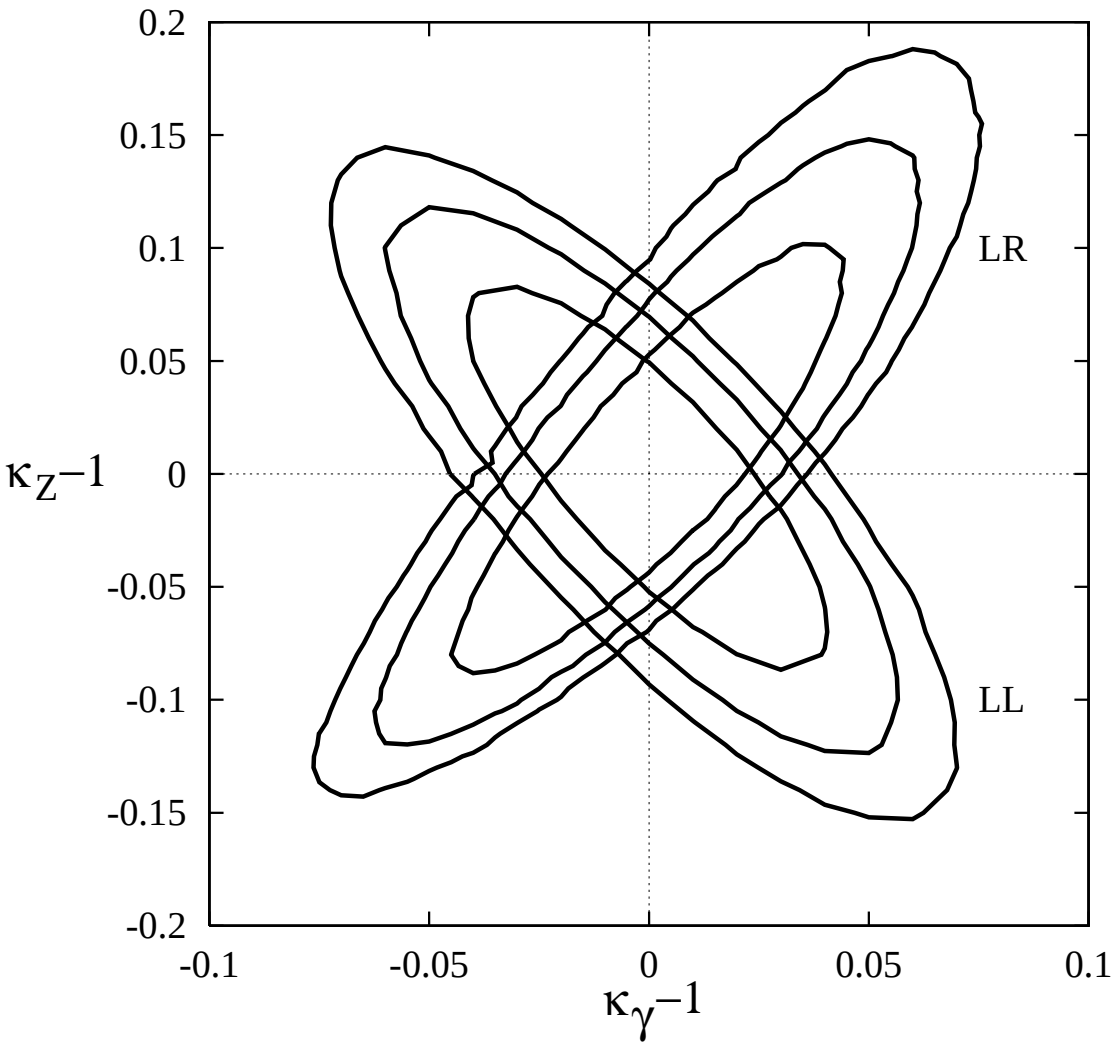

Figure 14: Contours of detectability at different confidence levels ($\chi^2 = 2, 4, 6$) of $\kappa_\gamma$ and $\kappa_Z$ for $LL$ and $LR$ polarizations. The remaining couplings assume their standard model values.

## 9 Conclusion

To summarize, we have described a few processes predicted by extensions of the standard model, which may be observed at an $e^-e^-$ facility. Some of these reactions are outstanding, in the sense that no other experiment can perform as well (*e.g.*, the search for Majorana neutrinos or the identification of $F = 0$ and $F = 2$ leptoquarks). Other reactions deliver informations which are comparable and complementary to those obtainable in other collider modes (*e.g.*, $Z'$ searches or anomalous gauge couplings).

Considering the potential benefits and the relative ease with which a positron beam can be replaced by an electron beam at a linear collider, $e^-e^-$ collisions emerge as a worthy endeavour for the high energy physics community.

## References

This list of references is by no means exhaustive. However, most of the cited papers have been published recently and contain the references to the relevant earlier publications. A more and nearly complete $e^-e^-$ bibliography can be accessed on the WWW at `http://pss058.psi.ch/cuypers/e-e-.html`.

1. C.A. Heusch, Proc. of the *1st Arctic Workshop on Future Physics and Accelerators*, Saariselka, 1994, SCIPP-95-02.
2. D. Choudhury, F. Cuypers, A. Leike, *Phys. Lett.* **B333** (1994) 531 [hep-ph/9404362].
3. *cf., e.g.,* F. Del Aguila *Acta Phys. Polon.* **B25** (1994) 1317 [hep-ph/9404323];
   S. Godfrey, *Phys. Rev.* **D51** (1995) 1402 [hep-ph/9411237].
4. P.H. Frampton, D. Ng, *Phys. Rev.* **D45** (1992) 4240;
   T.G. Rizzo, *Phys. Rev.* **D46** (1992) 910.
5. F. Cuypers, MPI-PHT-95-107, *Santa Cruz $e^-e^-$ Workshop*, Santa Cruz, CA, 4-5 Sep 1995 [hep-ph/9510443].
6. W. Buchmüller, R. Rückl, D. Wyler *Phys. Lett.* **B191** (1987) 442.
7. *cf., e.g.,* C.A. Heusch, P. Minkowski, *Nucl. Phys.* **B416** (1994) 3;
   G. Bélanger, F. Boudjema, D. London, H. Nadeau, ENSLAPP-A-537-95 [hep-ph/9508317].
8. P. Minkowski, *Prerequired Neutrino Properties for Observable Lepton Flavor Violation Effects at NLC*, these proceedings; C.A. Heusch, *Heavy Neutrino Search from $e^-e^-$ Collision*, these proceedings.
9. F. Cuypers, K. Kołodziej, R. Rückl, *Phys. Lett.* **B325** (1994) 243; *Nucl. Phys.* **B430** (1994) 231 [hep-ph/9405421].
10. F. Cuypers, G.J. van Oldenborgh, R. Rückl, *Nucl. Phys.* **B409** (1993) 128 [hep-ph/9305287].
11. F. Cuypers, *Phys. At. Nucl.* **56** (1993) 1460; *Yad. Fiz.* **56** (1993) 23.
12. H.P. Nilles, *Phys. Rep.* **110** (1984) 1;
    H.E. Haber, G.L. Kane, *Phys. Rep.* **117** (1985) 75.
13. *Selectron Searches in $e^-e^-$, $\gamma\gamma$ and $\gamma\gamma$ Scattering*, these proceedings.
14. K.I. Hikasa, *Phys. Lett.* **B164** (1985) 385; *erratum, ibid.* **B195** (1987) 623.
15. T.G. Rizzo, SLAC-PUB-95-7031, *Santa Cruz $e^-e^-$ Workshop*, Santa Cruz, CA, 4-5 Sep 1995 [hep-ph/9510296].
16. J.F. Gunion, UCD-95-36, *Santa Cruz $e^-e^-$ Workshop*, Santa Cruz, CA, 4-5 Sep 1995 [hep-ph/9510350].
17. D. Choudhury, F. Cuypers, *Nucl. Phys.* **B429** (1994) 33 [hep-ph/9405212].
18. F. Cuypers, K. Kołodziej, *Phys. Lett.* **B344** (1995) 365 [hep-ph/9411277].
19. F. Boudjema, Proc. of the Workshop *$e^+e^-$ Collisions at 500 GeV: the Physics Potential*, Munich/Annecy/Hamburg, 1992-93, DESY 93-123C, p. 177, Ed. P. Zerwas, and references therein.
20. *Manifestations of Strong Electroweak Symmetry Breaking in $e^-e^-$ Scattering*, these proceedings.